\documentclass[10pt]{article}
\usepackage{latexsym}
\usepackage[dvips]{color}
\usepackage{graphicx}
\usepackage{amsfonts}
\usepackage{amsmath}
\usepackage{amssymb}
\usepackage{psfrag}
\usepackage{hyperref}
\newtheorem{theorem}{Theorem}[section]

\newtheorem{proposition}[theorem]{Proposition}

\newtheorem{remarks}[theorem]{Remarks}
\newtheorem{remark}[theorem]{Remark}

\newcommand{\beq}{\begin{equation}}
\newcommand{\feq}[1]{\label{#1} \end{equation}}
\newcommand{\beqr}{\begin{eqnarray}}
\newcommand{\feqr}{\end{eqnarray}}
\def\non{\nonumber}

\newcommand{\rf}[1]{(\ref{#1})}
\begin{document}
\title{\bf The Partition Function of the Dirichlet Operator $\mathcal{D}_{2s}=\sum_{i=1}^{d}(-\partial_i^2)^{s}$ on a d-Dimensional Rectangle Cavity\unboldmath}
\author{Agapitos N. Hatzinikitas, \\
University of Aegean, 
School of Sciences, \\
Department of Mathematics, 
Karlovasi 83200,
Samos, Greece \\
Email: ahatz@aegean.gr}
\date{}
%%%%%%%%%%%%%%%%%%%%%%%%%%%%%%%%%%%%%%%%%%%%%%%%%%%%%%%%%%%%%%%%%%%%%%%%%%%%%%%%%%%%%%%%%%%%%%%%%%
%%%%%%%%%%%%%%%%%%%%%%%%%%%%%%%%%%%%%%%%%%%%%%%%%%%%%%%%%%%%%%%%%%%%%%%%%%%%%%%%%%%%%%%%%%%%%%%%%%
\maketitle
\begin{abstract}
In this letter we study the asymptotic behavior of the free partition function in the $t\rightarrow 0^+$ limit for a stochastic process which consists of $d-$independent, one-dimensional, symmetric, $2s-$stable processes in a hyperrectangular cavity $K \subset \mathbb {R}^d$ with an absorbing boundary. Each term of the partition function for this polyhedron  in d-dimensions can be represented by a quermassintegral and the geometrical information inherited by the eigenvalues for this solvable model is complete. We also demonstrate the correctness of our result by applying the method of images in one dimension. 
 \end{abstract}
%%%%%%%%%%%%%%%%%%%%%%%%%%%%%%%%%%%%%%%%%%%%%%%%%%%%%%%%%%%%%%%%%%%%%%%%%%%%%%%%%%%%%%%%%%%%%%%%%%%%%%%%%%%%%
%%%%%%%%%%%%%%%%%%%%%%%%%%%%%%%%%%%%%%%%%%%%%%%%%%%%%%%%%%%%%%%%%%%%%%%%%%%%%%%%%%%%%%%%%%%%%%%%%%%%%%%%%%%%%
\vspace{1cm}
%\newpage
%%%%%%%%%%%%%%%%%%%%%%%%%%%%%%%%%%%%%%%%%%%%%%%%%%%%%%%%%%%%%%%%%%%%%%%%%%%%%%%%%%%%%%%%%%%%%%%%%%%%%%%%%%%
\section{Introduction}

Trace formulas for heat kernels of  the fractional Laplacian $(-\Delta)^s,\, s\in (0,1)$ \cite{FG} and its Schr\"{o}dinger perturbations in spectral theory \cite{BY} have attracted a lot of attention recently due to the numerous applications to the mathematical physics, mathematical biology and finance. From a probabilistic point of view the fractional Laplacian on a domain $K\subset \mathbb{R}^d$, is a non-local operator which arises as the generator of a pure jump L\'{e}vy process killed upon exiting $K$. 
\par In a recent paper \cite{AH} we investigated the distribution of eigenvalues of the Dirichlet pseudo-differential operator $\sum_{i=1}^{d}(-\partial_i^2)^{s}, \, s\in (0,1)$, on an open and bounded subdomain  $K \subset \mathbb{R}^d$, which is defined by the principal value integral 
\begin{align}
& (- \partial_i^2)^s \psi_i(x_i) = C_{d=1,s} \, \textrm{P.V.} \int_{\mathbb{R}} \frac{\psi_i(x_i)-\psi_i(y_i)}{|x_i-y_i|^{1+2s}} dx_i
= C_{d=1,s} \lim_{\epsilon \rightarrow 0^+}\int_{|x_i-y_i| > \epsilon} \frac{\psi_i(x_i)-\psi_i(y_i)}{|x_i-y_i|^{1+2s}} dx_i, \non \\
& C_{d=1,s}=-2\textrm{cos}(\pi s)\Gamma(-2s).
\label{sec0 : eq1}
\end{align}
The $\psi_i(x_i)$ are restrictions of functions that belong to the fractional Sobolev space \cite{NPV}, $H^s(\mathbb{R})=W^{s,2}(\mathbb{R}), \, s\in (0,1)$, given by 
\begin{align}
H^s(\mathbb{R})\!=\!\{\psi_i \! \in \! L^2(\mathbb{R})\! : \! \frac{|\psi_i(x_i)-\psi_i(y_i)|}{|x_i-y_i|^{\frac{1}{2}+s}} \! \in \! L^2(\mathbb{R}\times \mathbb{R})\}
\label{sec0 : eq2}
\end{align}
and satisfy $\psi(x)=\prod_{i=1}^d\psi_i(x_i)\equiv 0$ in $\mathbb{R}^d \setminus \bar{K}$. We have also  predicted bounds on the sum of the first $N$ eigenvalues, the counting function, the Riesz means and the first-term asymptotic expansion of the partition function which was found to be  
\begin{align}
Z(t)=\int_{0}^{\infty}e^{-\mathcal{E}t}d\mathcal{N}(\mathcal{E})
=\frac{1}{(2\pi)^d}|\Omega|\frac{(2\Gamma(1+\frac{1}{2s}))^d}{t^{\frac{d}{2s}}}+o(t^{-\frac{d}{2s}})
\label{sec0 :eq3}
\end{align}
where the counting function $\mathcal{N}(\cdot)$ is\footnote{We denote by $\left\|x \right\|^{2s}=\sum_{i=1}^d |x_i|^{2s}$ the $2s$-norm and by $\left\|\cdot \right\|_2$ the Euclidean norm.} 
\begin{align}
\mathcal{N}(\mathcal{E}) =\sharp \left\{ n\in {\mathbb Z^d_+}: \left\| n \right\|^{2s}\leq \left(\frac{L}{\pi}\right)^{2s} \mathcal{E} (1+o(1))\right\}.
\label{sec0 : eq4} 
\end{align}
\par The present paper extends result \rf{sec0 :eq3} to all orders in $t^{-\beta}, \, \beta>0$, in the small time limit, and provides a novel result for a polyhedron's partition function which was lacking from the literature even for the ordinary Laplacian. Before stating our result, we review basic notions of convex geometry and apply them to a d-dimensional hyperrectangular parallelepiped in  Euclidean space. The explicit knowledge of the Dirichlet eigenvalues \rf{sec2 : eq1ad} as well as the Euler-Maclaurin summation formula \rf{sec2 : eq1} enable us to determine the asymptotic behavior of the partition function \rf{sec2 : eq7}. It is expressed in terms of the volume of $K$ and the rth quermassintegral. We also present an alternative proof by applying the method of images to the diffusion equation \rf{sec2 : eq15} with initial and boundary conditions.       
%%%%%%%%%%%%%%%%%%%%%%%%%%%%%%%%%%%%%%%%%%%%%%%%%%%%%%%%%%%%%%%%%%%%%%%%%%%%%%%%%%%%%%%%%%%%%%%%%%%%%%%%%%%
\section{Geometric Preliminaries of Convex Bodies in $\mathbb{R}^d$}
Let $C(\mathbb{K})$ be the family of all convex bodies (nonempty, compact, convex sets) in the d-dimensional Euclidean vector space. We denote by $B^{d}=\{x\in \mathbb{R}^d : \left\|x \right\|_2\leq 1\}$ the unit ball and $\mathbb{S}^{d-1}=\{x\in \mathbb{R}^d : \left\|x \right\|_2= 1\}$ the unit sphere. The Lebesgue measure on $\mathbb{R}^d$ is denoted by $\lambda_d$ and the spherical Lebesgue measure on $\mathbb{S}^{d-1}$ by $\sigma_{d-1}$. In particular we have: $\omega_d:=\lambda_d(B^d)=\pi^{d/2}/\Gamma(1+d/2)$ and $O_{d-1}:=\sigma_{d-1}(\mathbb{S}^{d-1})=d\omega_d=2\pi^{d/2}/\Gamma(d/2)$. If $K\in C(\mathbb{K})$ then the parallel body $K_{\rho}$ at distance $\rho>0$ is given by 
\begin{align}
K_{\rho}:=\{x\in \mathbb{R}^d : \textrm{dist.}(x,K)\leq \rho\} 
\end{align}
where $\textrm{dist.}(\cdot,K)$ denotes the Euclidean distance from $K$. According to Steiner's formula the volume of $K_{\rho}$ is given by
\begin{align}
\lambda_{\rho}(K_{\rho})=\sum_{m=0}^{d} C^m_d W_m(K) \rho^m
=\sum_{j=0}^{d} \omega_{d-j} V_j(K) \rho^{d-j}, \, \rho \geq 0, \quad C^m_d=\left( \!\! \begin{array}{c} d \\ m\end{array} \!\! \right)
\label{sec1 : eq1}
\end{align}
where $W_m : C(\mathbb{K})\rightarrow \mathbb{R}$ is the mth \textit{quermassintegrale}, or mean cross-sectional measure, introduced by Minkowski, and $V_j : C(\mathbb{K})\rightarrow \mathbb{R}$ is the jth intrinsic volume\footnote{The word intrinsic here means that the quantity under study does not depend on the dimension of the ambient space.}. The $W_m$ in \rf{sec1 : eq1} is defined by \cite{LS}
\begin{align}
W_m(K)=\frac{(d-m) O_{m-1}\cdots O_0}{d O_{d-2}\cdots O_{d-m-1}} \!\! \int_{G_{m,d-m}} \!\!\!\!\!\!\!\!\!\!\! V(K'_{d-m})dL_{m [O]}
\label{sec1 : eq2}
\end{align}
where  $V(K'_{d-m})$ denotes the volume of the convex set of all intersection points of the $(d-m)$-plane, passing through the fixed point $O$, with the $m-$planes orthogonal to it and $dL_{m [O]}$ is the invariant volume element of the Grassmannian manifold $G_{m,d-m}$ which is the set of $m-$dimensional planes in $\mathbb{R}^{d-m}$. Setting $m=d-1$ into \rf{sec1 : eq2} one can define another functional, the so-called mean breadth of $K$, as follows
\begin{align}
\bar{b}= \frac{2}{O_{d-1}} \int_{G_{d-1,1}}\!\!\!\!\!\!\!\!\!V(K'_{1})dL_{d-1 [O]} 
= \frac{2}{\omega_d}W_{d-1}(K) 
\label{sec1 : eq2a}.
\end{align}   
\par If the boundary $\partial K$ of a convex set is a hypersurface $\Sigma$ of class $C^2$, the quermassintegrals $W_m$ can be expressed by means of the integrals of mean curvature of $\partial K$. The mth integral of mean curvature $M_m(\Sigma)$ is defined by
\begin{align}
M_m(\Sigma)=\frac{1}{C^m_{d-1}}\int_{\Sigma} S^{(d-1)}_m(\kappa) dA
\label{sec1 : eq3}
\end{align}
where $S^{(d-1)}_m(\kappa)=\sum_{1\leq i_1< \cdots < i_{m}\leq d-1} \kappa_{i_1}\cdots \kappa_{i_m}$ is the mth-elementary symmetric function of the $(d-1)$ principal curvatures and $dA$ is the area element of $\Sigma$. The volume of the parallel body $K_{\rho}$ can be written then as
\begin{align}
\lambda_{\rho}(K_{\rho})=V(K)+\sum_{m=0}^{d-1}\frac{C^m_{d-1}}{m+1} M_m(\partial K) \rho^{m+1}
\label{sec1 : eq4}
\end{align}
and by comparison with \rf{sec1 : eq1} we end up with
\begin{align}
M_m(\partial K)=d W_{m+1}(K).
\label{sec1 : eq5}
\end{align}
This relation is well defined as long as $\partial K$ is $C^2$. If $K$ does not have a smooth boundary then we compute 
\begin{align}
\lim_{\rho \rightarrow 0^+}M_m(\partial K_{\rho})=d W_{m+1}(K).
\label{sec1 : eq6}
\end{align}
\par In the present article we focused on the hyperrectangular parallelepiped with edges $a_j, \, j=1,\cdots,d$. The mth intrinsic volume and quermassintegral are given by
\begin{align}
V_m(K)&= S^{(d)}_m (a), V_0(a)=1 \label{sec1 : eq6a} \\
W_m(K)&= \frac{\omega_m}{C^m_{d-1}} V_{d-m}(K). \label{sec1 : eq6b}
\end{align}
The mean breadth of $K$ can be computed by combining \rf{sec1 : eq3} with \rf{sec1 : eq2a} for the parallel body $K_{\rho}$ and taking the $\rho \rightarrow 0^+$ limit in the end. A hyperrectangular parallelepiped has a total of $2d$ faces, $2^d$ vertices and $2^{d-1} d$ edges. The mean curvature is defined by 
\begin{align}
H_{d-2}(R)=\frac{1}{d-1}S^{(d-1)}_{d-2}(R) 
=\frac{1}{d-1}\sum_{1\leq i_1< \cdots < i_{d-2}\leq d-1} \frac{1}{R_{i_1}\cdots R_{i_{d-2}}}
\label{sec1 : eq7}
\end{align}
where $R_i$ are the principal radii of curvature of $\partial K_{\rho}$. The boundary of $\partial K_{\rho}$ consists of hyperplanes at the faces, hyperspheres at the vertices and hypercylinders at the edges of $K$. The values of $H$ for $\partial K_{\rho}$ are
\begin{align}
H_{d-2}(R)\!=\!\left\{\!\! \begin{array}{cl} 0, \!\!&\!\!\! \textrm{hyperplanes} \\ \!\! \frac{1}{\rho^{d-2}}, \!\!&\!\!\! \textrm{hyperspheres} \\ \!\! \frac{1}{(d-1)\rho^{d-2}}, \!\!&\!\!\! \textrm{lateral face of hypercylinders}  \end{array}\right.
\label{sec1 : eq8}
\end{align}
A careful calculation gives
\begin{align}
W_{d-1}(K)=\lim_{\rho \rightarrow 0^+} W_{d-1}(\partial K_{\rho})= \frac{1}{d}\omega_{d-1}\sum_{i=1}^da_i
\label{sec1 : eq9}
\end{align}
and therefore
\begin{align}
\bar{b}=\frac{2}{d}\frac{\omega_{d-1}}{\omega_{d}}V_1(K).
\label{sec1 : eq10}
\end{align}  
%%%%%%%%%%%%%%%%%%%%%%%%%%%%%%%%%%%%%%%%%%%%%%%%%%%%%%%%%%%%%%%%%%%%%%%%%%%%%%%%%%%%%%%%%%%%%%%%%%%%%%%%%%%
\section{The Partition Function for the Operator $\mathcal{D}_{2s}$ on $K$}

\par Let $\textbf{X}(t)=\{X_i(t)\}_{i=1}^d, \, t>0$ be a collection of independent, one-dimensional, symmetric $2s-$stable processes in $\mathbb{R}$ and denote by $\{P_K(t)\}_{t\geq 0}$ the semi group on $L^2(K)$ of $\textbf{X}(t)$ killed upon exiting $K$. Its transition density $p_{K}(x,t; y)$ satisfies
\begin{align}
P_K(t) f(x)=\int_{K}p_{K}(x,t; y)f(y)dy.
\label{sec2 : eq1fa}
\end{align}
In \cite{KWA} and \cite{AH}, performing a slight modification, it has been proved the following proposition:
\begin{proposition}
On the open and bounded hyperrectangular parallelepiped $K\subset {\mathbb R^d}$ of side lengths $a_i, \, i=1,\cdots,d$, the eigenvalues for the homogeneous Dirichlet problem
\begin{align}
\left(\sum_{i=1}^{d}(-\partial_i^2)^s \psi_n \right)(x) &=\mathcal{E}_n(s)\psi_n(x), \,\, \textrm{in} \,\, K; \,\,\mathcal{E}_n=\frac{E_n}{D_{2s}} \non \\
 \psi_n(x)&= 0 \,\,  \textrm{on} \,\, \mathbb{R}^d \setminus \bar{K}
\label{sec2 : eq1ac}
\end{align}
are given by 
\begin{align}
\mathcal{E}_{n}(s) =\sum_{i=1}^d \left| \frac{n_i \pi}{a_i}-\frac{(1-s)\pi}{2a_i}\right|^{2s}\!\!\!+\sum_{i=1}^{d}O(\frac{1}{n_i}), \, \, n\in {\mathbb Z_+^d} 
\label{sec2 : eq1ad}
\end{align}
 where $\{\psi_n\}_{n=1}^{\infty}$ forms an orthonormal basis in $L^2(K)$ with $\psi_n(x)=c_n\prod_{j=1}^{d}\psi_{n_j}(x_j)$, none of the indices $n_j$ vanishes, and $D_{2s}=(\hbar^2/[M])^s$ is a constant with dimensions $[D_{2s}]=[M]^s [L]^{4s}/[T]^{2s}$.
\end{proposition}
Arranging the positive, real and discrete spectrum of $\mathcal{D}_{2s}$ in increasing order (including multiplicities), we have 
\begin{align}
& 0<\mathcal{E}_1(K) < \mathcal{E}_2(K) \leq \mathcal{E}_3(K) \leq \cdots \non \\
& \textrm{and} \,\, \lim_{n\rightarrow \infty} \mathcal{E}_n(K) =\infty.
\label{sec2 : eq1ae}
\end{align}
The simplicity of $\mathcal{E}_{n_j}$ is conjectured to hold for $s\in (0,1)$, proved for $s=1/2$ and $s\in [1/2,1)$ in \cite{KKM}, \cite{KWA} respectively.
\par Taking into account the previous proposition the generator of the semi group acts on the orthonormal set $\psi_n$ as
\begin{align}
e^{-t \mathcal{D}_{2s}|_{K}} \psi_n(x)=e^{-t \mathcal{E}_{n}} \psi_n(x), \, x\in K, \, t>0.
\label{sec2 : eq1fb}
\end{align}
The transition density is then given by
\begin{align}
p_{K}(x,t;y)=\sum_{n\in \mathbb{Z}^d_+}e^{-t\mathcal{E}_n} \bar{\psi}_n(x) \psi_n(y)
\label{sec2 : eq1fc}
\end{align}   
and the partition function (or trace of the heat kernel), using \rf{sec2 : eq1fc}, satisfies
\begin{align}
Z_{K}(t,s)&\equiv \textrm{Tr}(e^{-t\mathcal{D}_{2s}})=\int_{K} p_K(x,t;x)dx 
=\sum_{n\in \mathbb{Z}^d_+} e^{-t\mathcal{E}_n} \int_K |\psi_n(x)|^2 dx \non \\
&=\prod_{j=1}^d\left(\sum_{n_j\geq 1}e^{-t\mathcal{E}_{n_j}}\right)=\prod_{j=1}^d Z_j(t,s), \, t>0.
\label{sec2 : eq1af}
\end{align}
Thus from \rf{sec2 : eq1af} we observe that the total partition function is written as the disjoint product of partition functions for each spatial dimension.  
The sum $\sum_n exp(-t \mathcal{E}_n)$ will be calculated utilizing the Euler-Maclaurin summation formula \cite{AAR} which states that:
\begin{theorem}
Suppose f is a decreasing function with continuous derivatives up to order p. Then 
\begin{align}
\sum_{k=n+1}^{m}f(k)=\int_n^m f(u) du 
+\sum_{l=1}^{p} (-1)^l \frac{B_l(0)}{l!}\{f^{(l-1)}(m)-f^{(l-1)}(n)\} 
+ \frac{(-1)^{p-1}}{p!} \int_n^m B_p(u-[u]) f^{(p)}(u) du,
\label{sec2 : eq1}
\end{align}
where $B_p(u)$ are the Bernoulli polynomials and $B_l(0)$, the Bernoulli numbers. 
\end{theorem}
Applying \rf{sec2 : eq1} in one dimension, in the zero time limit, we find for $c(s)=\frac{1-s}{2}$
\begin{align}
\sum_{n_j=1}^{\infty}e^{-\left[\frac{\pi}{a_j}(n_j-c)\right]^{2s}t} \stackrel{\textrm{as} \, t\rightarrow 0^+}{\sim}
\frac{a_j}{\pi t^{\frac{1}{2s}}}\Gamma\left(1+\frac{1}{2s}\right)-\frac{s}{2}, \,\, s\in (0,1].                                                                                     
\label{sec2 : eq5}
\end{align}
\begin{remarks}
\begin{enumerate}
\item The functions $f(t,u,s)$ belong to the following spaces
\begin{align}
f \in \left\{ \begin{array}{l} C^1_u((0,\infty)^2\times (\frac{1}{2},1) ) \\ C^{\infty}_u((0,\infty)^2\times \{\frac{1}{2}\}  \\ C^0((0,\infty)^2\times (0,\frac{1}{2})) \end{array}\right. 
\label{sec2 : eq5a} 
\end{align}
where the subscript $u$ indicates that the functions are differentiable w.r.t. $u$.
\item The case $s=\frac{1}{2}$ can be summed exactly since it turns out to be a progression. The result is
\begin{align}
Z_j(t,\frac{1}{2})=\frac{\left(\cosh(\frac{\pi}{4a_j}t)-\sinh(\frac{\pi}{4a_j}t)\right)}{\sinh(\frac{\pi}{2a_j}t)} 
\stackrel{\textrm{as} \, t\rightarrow 0^+}{\sim} \frac{a_j}{\pi t}-\frac{1}{4}.
\label{sec2 : eq6}
\end{align} 
\end{enumerate}
\end{remarks}
Substituting \rf{sec2 : eq5} into \rf{sec2 : eq1af}, for $s\in [1/2,1)$ we obtain
\begin{align}
Z_{K}(t,s) & \stackrel{\textrm{as} \, t\rightarrow 0^+}{\sim}\prod_{j=1}^d\left( \frac{a_j}{\pi t^{\frac{1}{2s}}}\Gamma\left(1+\frac{1}{2s}\right)-\frac{s}{2}\right) 
=\sum_{m=0}^{d}\left(\frac{1}{\pi t^{\frac{1}{2s}}} \Gamma\left(1+\frac{1}{2s}\right)\right)^{m}\!\!\! \left(-\frac{s}{2}\right)^{d-m} \!\!\! V_m(K) \non \\
&= \left(\frac{1}{\pi t^{\frac{1}{2s}}} \Gamma \!\! \left(\!\! 1\!+\! \frac{1}{2s}\!\! \right)\!\! \right)^d \!\!\! V_d(K) \!\! 
+\sum_{r=0}^{d-1}\left(\frac{1}{\pi t^{\frac{1}{2s}}} \Gamma \left(1+\frac{1}{2s}\right)\right)^{d-r} 
\times \left(-\frac{s}{2}\right)^{r}  \frac{d-r}{O_{d-r-1}}\left(\begin{array}{c} d\\ d-r \end{array} \right)W_{d-r}(K) 
\label{sec2 : eq7}
\end{align}
where $V_m(K)$, $W_{d-r}(K)$ are given by \rf{sec1 : eq6a} and \rf{sec1 : eq6b}. \\
\begin{remarks}
\begin{enumerate}
\item The term $V_d(K)=\prod_{j=1}^d a_{j}$ is the d-dimensional volume of the hyperrectangle parallelepiped and $V_1(K)$ can be written in terms of the mean breadth of $K$ as
\begin{align}
V_1(K)=\frac{d \omega_d}{2\omega_{d-1}}\bar{b}
\label{sec2 : eq71a}
\end{align}
using \rf{sec1 : eq10}. 
\item In two dimensions the trace of the heat kernel for the ordinary Laplacian ($s=1$), provided that the domain is a rectangle of sides $a_1, a_2$, is given by 
\begin{align}
Z_K(1,t)=\frac{1}{4}\left[\Theta\left(\frac{\pi t}{a_1^2}\right)-1\right]\left[\Theta\left(\frac{\pi t}{a_2^2}\right)-1\right]
\label{sec2 : eq11}
\end{align}
where $\Theta(x)$ is the familiar Riemann theta function\footnote{The Jacobi theta function is defined by 
\begin{displaymath} \vartheta \left[\begin{array}{c} a \\b \end{array}\right](z,\tau)=\sum_{n\in \mathbb{Z}} e^{i\pi (n+a)^2 \tau+ 2i \pi (n+a) (z+b)}.\end{displaymath}},
\begin{align}
\Theta(x)=\vartheta \left[\begin{array}{c} 0 \\0 \end{array}\right](0,-ix)=\sum_{n\in \mathbb{Z}} e^{-\pi n^2 x}.
\label{sec2 : eq11a}
\end{align}  
The Poisson summation formula\footnote{The Poisson formula states \begin{displaymath} \sum_{n\in \mathbb{Z}} e^{-\pi n^2 A+2\pi n s A}=\frac{1}{\sqrt{A}}e^{\pi s^2 A}\sum_{m\in \mathbb{Z}} e^{-\pi m^2 A^{-1}-2i\pi m s}\end{displaymath} and can be proved by using $\sum_m e^{2i\pi r m}=\sum_n \delta(r-n)$.} can be casted into the form \cite{CH}
\beqr
\Theta(x)=\frac{1}{\sqrt{x}}\Theta(\frac{1}{x})
\label{sec2 : eq11b}
\feqr
and utilized to determine the asymptotic behavior of the partition function in the $t\rightarrow 0^+$ limit. Therefore we reproduce the well-known result
\begin{align}
Z_{K}(t,s=1)\sim \frac{1}{4\pi t}V_2(K)-\frac{1}{8\sqrt{\pi t}}V_1(K)+\frac{1}{4}
\label{sec2 : eq12}
\end{align}
where $V_2(K)$ is the surface area of $K$ and $V_1(K)$ is the length of the boundary $\partial K$. An independent calculation using \rf{sec2 : eq7} gives identical result. Kac, in \cite{K}, extended the dimensionless corner correction for a closed polygon with obtuse angles through a complicated integral. Later it was reported in \cite{MS} that D.B.~Ray obtained the correction  
\begin{align}
\sum_{i=1}^n \frac{\pi^2- \phi_i^2}{24 \pi \phi_i}, \, 0<\phi_i< 2\pi
\label{sec2 : eq13}
\end{align}
for arbitrary angles by expressing the Green's function as a Kontorovich-Lebedev transform. It is noteworthy that if we approximate a circle by an inscribed regular polygon then \rf{sec2 : eq13} becomes
\begin{align}
\sum_{i=1}^n \frac{\pi^2- \phi_i^2}{24 \pi \phi_i}=\frac{1}{6} \left(\frac{n-1}{n-2}\right).
\label{sec2 : eq13a}
\end{align}
In \rf{sec2 : eq13a} as the number of edges of the polygon tends to infinity the sum converges to the topological invariant constant $1/6$. For simply connected, open with compact boundary two-dimensional Riemannian manifolds $(\mathcal{M},g)$, where $g$ is the metric tensor, the partition function contains the $t$-independent term given by \cite{MS}
\begin{align}
\frac{E}{6}=\frac{1}{12 \pi}\int_{\mathcal{M}} Ric \sqrt{\det g} dx
\label{sec2 : eq14}
\end{align}
where $E$ is the Euler characteristic and $Ric$ is the Ricci scalar of the manifold $\mathcal{M}$. 
\end{enumerate}
\end{remarks}
%%%%%%%%%%%%%%%%%%%%%%%%%%%%%%%%%%%%%%%%%%%%%%%%%%%%%%%%%%%%%%%%%%%%%%%%%%%%%%%%%%%%%%%%%%%%%%%%%%%%%%%%%%%%%
%%%%%%%%%%%%%%%%%%%%%%%%%%%%%%%%%%%%%%%%%%%%%%%%%%%%%%%%%%%%%%%%%%%%%%%%%%%%%%%%%%%%%%%%%%%%%%%%%%%%%%%%%%%%%
\section{An Alternative Approach Based on the Fractional Diffusion Equation}

\par The transition function $p_j$ (or elementary solution) is the solution of the following diffusion problem in $K_j=(0,a_j)\subset \mathbb{R}$ described by the equation
\begin{align}
\frac{\partial p_{j}(x,t;y)}{\partial t}=\left(-\frac{d^2}{dx^2}\right)^s p_{j}(x,t;y), \,\, x,y \in K_j, \, t>0 \label{sec2 : eq15}
\end{align}
with initial and boundary conditions
\begin{subequations}
\begin{align}
\int_{K_j}\lim_{t\rightarrow 0^+} p_{j}(x,t;y) dx&=\int_{K_j} \delta(|x-y|)dx=1, \tag{38.a} \\
\lim_{x\rightarrow q\in \partial K_j}p_{j}(x,t;y)&=0 \tag{38.b}. 
\end{align}
\end{subequations}
The initial condition expresses the fact that a point source is located at $x=y$ and described by a Dirac-$\delta$ generalized function while the boundary condition implies that the process is killed on reaching the boundary $\partial K_j$. The solution of \rf{sec2 : eq15} with the initial condition (38.a) on $\mathbb{R}$ is given by 
\begin{align}
 p_{j,\mathbb{R}}(x,t;y)=p_{j,\mathbb{R}}(|x-y|,t) 
=\frac{1}{2\pi}\int_{\mathbb{R}} e^{-t|k-\textrm{sign}(k)\frac{\pi g(s)}{a_j}|^{2s}+i(k-\textrm{sign}(k)\frac{\pi g(s)}{a_j})|x-y|} dk, \,\, g(s)=\frac{1-s}{2}
\label{sec2 : eq16}
\end{align}
where the appearance of the function $\textrm{sign}(k)\frac{\pi g(s)}{a_j}$ will be apparent shortly. Note that due to the translational invariance of the integral it can be absorbed in $k$ and written in the usual way. 
\par The problem in $K$ can be solved using the familiar method of images according to which we produce, for the point source at $x=y$, two infinite sequences of images, with alternating strengths $\pm 1$, by successive reflections w.r.t. the boundary points $x=0$ and $x=a_j$. In this way we obtain
\begin{align}
\begin{array}{ccccc}
\textrm{strength}:  & \!\!\!  -1, & \!\!\! +1, & \!\!\! -1, & \!\!\!\! \cdot \\ 
\textrm{starting from x=0}: & \!\!\! -y, & \!\!\! y+2a_j, & \!\!\! -y-2a_j, & \!\!\!\! \cdot \\
\textrm{starting from x=a}: & \!\!\! -y+2a_j, & \!\!\! y-2a_j, & \!\!\! -y+4a_j, & \!\!\!\! \cdot
\end{array}
\end{align}
The solution is then written as
\begin{align}
p_{j}(|x-y|,t)=\sum_{n\in \mathbb{Z}} p_{j,\mathbb{R}}(|x-y-2na_j|,t) 
-\sum_{n\in \mathbb{Z}}p_{j,\mathbb{R}}(|x+y-2na_j|).
\label{sec2 : eq17}
\end{align}
One can check that both the  initial and boundary conditions are fulfilled. The $x\rightarrow y$ limit of \rf{sec2 : eq17} gives
\begin{align}
\lim_{x\rightarrow y}p_{j}(|x-y|,t)=p_{j}(0,t) 
=p_{j,\mathbb{R}}(0,t)+2\sum_{n\in \mathbb{N}\setminus \{0\}} p_{j,\mathbb{R}}(|2na_j|,t)  
-\sum_{n\in \mathbb{Z}}p_{j,\mathbb{R}}(|2(y-na_j)|,t)
\label{sec2 : eq19}
\end{align}
and the contribution of each term to the partition function $Z_{j}(t)=\int_{K}p_{j}(y,t;y)dy$ is
\begin{align}
& \int_{0}^{a_j}p_{j,\mathbb{R}}(0,t)dy =\frac{a_j}{\pi t^{\frac{1}{2s}}}\Gamma\left(1+\frac{1}{2s} \right)+g(s)
\label{sec2 : eq20} \\
& 2\sum_{n\in \mathbb{N}\setminus \{0\}}\int_0^{a_j} \lim_{t \rightarrow 0^+} p_{j,\mathbb{R}}(|2na_j|,t)dy =\sum_{n \in \mathbb{N}\setminus \{0\}} \delta(n)=0
\label{sec2 : eq20a} \\
& \sum_{n\in \mathbb{Z}}\int_0^{a_j} \lim_{t \rightarrow 0^+} p_{j,\mathbb{R}}(|2(y-na_j)|,t)dy =\frac{1}{2}.
\label{sec2 : eq21}
\end{align}
Therefore
\begin{align}
Z_{j}(t)\stackrel{t\rightarrow 0^+}{\sim}\frac{a_j}{\pi t^{\frac{1}{2s}}}\Gamma\left(1+\frac{1}{2s} \right)+\left(-\frac{s}{2}\right)
\label{c}
\end{align}
as promised.
\begin{remark}
The diffusion initial-boundary value problem for $K=(0,a_1)\times (0,a_2)$ with $s=1$, can also be solved applying the method of images in a more general setting. One can construct the infinite sequence of image source points $y_{(n,m)}=(\pm y_1\pm 2na_1, \pm y_2\pm 2ma_2)$ with $n,m=0,1,2,\cdots$ by successive reflections on the opposite sides of the rectangle. The solution then reads
\begin{align}
p_{j}(\left\|x-y\right\|_2,t) 
=\sum_{n,m \in \mathbb{Z}}\left(p_{\mathbb{R}}(\left\|x-y_{++}\right\|_2,t)+p_{\mathbb{R}}(\left\|x-y_{--}\right\|_2,t)\right) 
- \sum_{n,m \in \mathbb{Z}}\left(p_{\mathbb{R}}(\left\|x-y_{+-}\right\|_2,t)+p_{\mathbb{R}}(\left\|x-y_{-+}\right\|_2,t)\right)
\label{sec2 : eq22}
\end{align}
where
\begin{align}
y_{++}& =(y_1+2na_1,y_2+2ma_2) \non \\
y_{--}&=(-y_1+2na_1,-y_2+2ma_2) \non \\
y_{+-}&=(y_1+2na_1,-y_2+2ma_2) \non \\
y_{-+}&=(-y_1+2na_1,y_2+2ma_2).
\label{sec2 : eq23}
\end{align}
\end{remark}
%%%%%%%%%%%%%%%%%%%%%%%%%%%%%%%%%%%%%%%%%%%%%%%%%%%%%%%%%%%%%%%%%%%%%%%%%%%%%%%%%%%%%%%%%%
%%%%%%%%%%%%%%%%%%%%%%%%%%%%%%%%%%%%%%%%%%%%%%%%%%%%%%%%%%%%%%%%%%%%%%%%%%%%%%%%%%%%%%%%%%%%%%%%%%%%%%%%%
\section{Further Speculations}
A remarkable consequence of the previous analysis is that the partition function is strongly related to the index (or characteristic exponent) $s$ of the symmetric L\'{e}vy flight. For the asymmetric case parameters such as the skewness, scale and location will also come into play. Importantly, relation \rf{sec2 : eq13} will also acquire a characteristic exponent $s-$dependence since the hyperrectangle calculation predicts the value $(-s/2)^d$. The fate of \rf{sec2 : eq14} requires further investigation. The basic obstacle is the lack of knowledge of the Laplace-Beltrami operator in the case of a symmetric L\'{e}vy flight.     
\par Another challenging open problem would be one to predict the partition function of an open manifold with convex and compact boundary knowing the sequence of polyhedral partition functions which approximates it. This issue is related to the geometric measure theory and fully resolving it will lead to the invention of new results in higher dimensions than those presented in \cite{MS}.     
%%%%%%%%%%%%%%%%%%%%%%%%%%%%%%%%%%%%%%%%%%%%%%%%%%%%%%%%%%%%%%%%%%%%%%%%%%%%%%%%%%%%%%%%%%
%%%%%%%%%%%%%%%%%%%%%%%%%%%%%%%%%%%%%%%%%%%%%%%%%%%%%%%%%%%%%%%%%%%%%%%%%%%%%%%%%%%%%%%%%%%%%%%%%%%%%%%%%

\end{document}